\documentclass[12pt]{article}
\usepackage{jheppub}
\usepackage{cite}
\usepackage{slashbox}

\usepackage[utf8]{inputenc} 
\usepackage[T1]{fontenc}    
\usepackage{hyperref}       
\usepackage{url}            
\usepackage{booktabs}       
\usepackage{nicefrac}       
\usepackage{microtype}      
\usepackage{enumitem}
\usepackage{amsmath}
\usepackage{wrapfig}
\usepackage{graphicx}
\usepackage{cancel}

\usepackage{tikz}
\usepackage{tikz-feynman}
\usepackage{comment}

\DeclareMathOperator{\tr}{tr}

\newcommand{\cL}{\mathcal{L}}
\newcommand{\cF}{\mathcal{F}}

\def\bea{\begin{eqnarray}}
\def\eea{\end{eqnarray}}

\def\be{\begin{equation}}
\def\ee{\end{equation}}

\def\eqn#1{eq.~\eqref{#1}}

\usepackage{todonotes}

\makeatletter
\DeclareRobustCommand*{\bfseries}{%
  \not@math@alphabet\bfseries\mathbf
  \fontseries\bfdefault\selectfont
  \boldmath
}

\makeatother

\title{Recurrent Features of Amplitudes in Planar \texorpdfstring{$\mathcal{N}=4$}{N=4} Super Yang-Mills Theory}

\author[\tt a]{Tianji Cai,}
\author[\tt b,c]{François Charton,}
\author[\tt d]{Kyle Cranmer,}
\author[\tt a]{Lance J.\ Dixon,}
\author[\tt d]{Garrett W.\ Merz,}
\author[\tt e,f]{Matthias Wilhelm}
\affiliation[\tt a]{SLAC National Accelerator Laboratory}
\affiliation[\tt b]{FAIR, Meta}
\affiliation[\tt c]{Ecole des Ponts}
\affiliation[\tt d]{Data Science Institute and Physics Department, University of Wisconsin-Madison}
\affiliation[\tt e]{Niels Bohr Institute, University of Copenhagen}
\affiliation[\tt f]{Institute for Mathematics and Computer Science, University of Southern Denmark}

\abstract{%
The planar three-gluon form factor for the chiral stress tensor operator in planar maximally supersymmetric Yang-Mills theory is an analog of the Higgs-to-three-gluon scattering amplitude in QCD.  The amplitude (symbol) bootstrap program has provided a wealth of high-loop perturbative data about this form factor, with results up to eight loops available.  The symbol of the form factor at $L$ loops is given by words of length $2L$ in six letters with associated integer coefficients.
In this paper, we analyze this data, describing patterns of zero coefficients and relations between coefficients. We find many sequences of words whose coefficients are given by closed-form expressions which we expect to be valid at any loop order.
Moreover, motivated by our previous machine-learning analysis, we identify simple recursion relations that relate the coefficient of a word to the coefficients of particular lower-loop words.
These results open an exciting door for understanding scattering amplitudes at all loop orders.
}

\begin{document}
\maketitle

\section{Introduction}

Scattering amplitudes are key quantities in Quantum Field Theory (QFT), and a crucial ingredient for precision prediction in contexts ranging from collider physics to gravitational-wave observations. In the study of scattering amplitudes, a prominent role is played by planar maximally supersymmetric Yang--Mills theory (planar $\mathcal{N}=4$ SYM theory); see e.g.~Ref.~\citep{Travaglini:2022uwo} for a review. Although it is not a realistic theory of nature, its simplicity has made it an ideal testing ground for developing techniques that are later applied to realistic theories such as quantum chromodynamics (QCD). Moreover, planar $\mathcal{N}=4$ SYM theory provides the opportunity to study properties of scattering amplitudes at very high loop orders, and even a chance of obtaining results at any loop order.

\begin{figure}[t]
    \centering
  \begin{minipage}[b]{0.25\textwidth}
    \includegraphics[width=\textwidth]{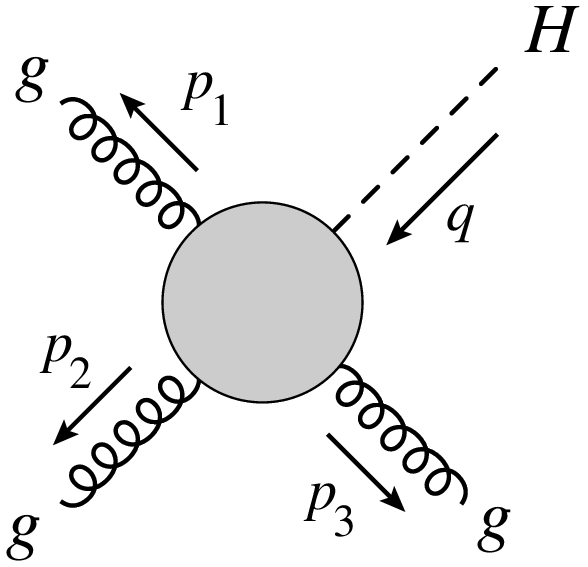}
  \end{minipage}
   \caption{The three-gluon form factor arising in the large-top-mass limit of a Higgs-to-three-gluon amplitude. In this paper, we consider its analog in planar $\mathcal{N}=4$ SYM theory.}
    \label{fig:three-gluon-form-factor}
\end{figure}

Probably the simplest non-trivial scattering amplitude in planar $\mathcal{N}=4$ SYM theory is the three-gluon form factor of the operator $\tr(F_+^2)$, where $\tr$ is a color trace and $F_+$ is the self-dual part of the field strength; see figure \ref{fig:three-gluon-form-factor}. This form factor is the analog of the Higgs-to-three-gluon amplitude in QCD in the limit of a large top quark mass \citep{Wilczek:1977zn,Shifman:1978zn,Gehrmann:2011aa,Brandhuber:2012vm}. Moreover, via the so-called antipodal duality, it is related to the (maximally helicity violating) six-gluon amplitude in planar $\mathcal{N}=4$ SYM theory \citep{Dixon:2021tdw}. 

The infrared-finite part of the three-gluon form factor, $\mathcal{F}$,  depends on only two dimensionless variables. It is conjectured to be expressible in terms of two-dimensional harmonic polylogarithms (2dHPLs) \citep{Gehrmann:2000zt} at any loop order \citep{Dixon:2020bbt,Dixon:2022rse}. 

Integrability (see Ref.~\citep{Beisert:2010jr} for a review) in the form of the form factor operator product expansion (FFOPE) \citep{Basso:2013vsa,Sever:2020jjx,Sever:2021nsq,Sever:2021xga} has given us an all-loop understanding of this form factor in an expansion around the collinear limit $p_1\,||\,p_2$.
The symbol bootstrap program \citep{Dixon:2011pw, Caron-Huot:2016owq,Caron-Huot:2019vjl} is complementary to the FFOPE; see Ref.~\citep{Caron-Huot:2020bkp} for a review. It allows one to obtain perturbative results -- so far up to eight-loop order \citep{Dixon:2020bbt,Dixon:2022rse} -- based on the symbol \citep{Goncharov:2010jf}, which is part of the Hopf algebra of multiple polylogarithms~\citep{Goncharov:2001iea}.
The symbol of the three-gluon form factor at $L$ loops is given by a sum of \emph{words}, i.e.\ sequences of $2L$ letters drawn from the alphabet $\{a,b,c,d,e,f\}$, multiplied by associated \emph{integer coefficients}.  The letters $\{a,b,c,d,e,f\}$ are simple functions of the dimensionless kinematical variables $u,v,w$, defined in eq.~\eqref{eq: definition abcdef} below.

In the symbol bootstrap program, the coefficients of the words are determined one loop order at a time via large systems of linear equations. These linear equations include homogeneous equations, which stem from mathematical consistency requirements and physical principles and are usually imposed first, as well as inhomogeneous equations encoding limiting behavior, including the FFOPE predictions of integrability, which are usually imposed afterwards.  At eight loops, the highest order currently calculated, the result for the symbol consists of $1.7$ billion sequences with non-vanishing coefficients.  The size of the system of linear equations constitutes a major bottleneck to bootstrapping higher loop orders, and calls for more efficient approaches.
    
In Ref.~\citep{Cai:2024znx}, we found that the relationship between words and their coefficients can be ``learned'' by machine-learning methods. In particular, we trained transformers \citep{Vaswani:2017lxt}, the architecture behind ChatGPT and other large-language models, to predict coefficients associated with words, after providing such models with many examples of coefficient-word pairings. Moreover, in what we called a \emph{strike-out} experiment, we found indications that the coefficients of many words at $L$ loops can be determined by coefficients of words at $L-1$ loops that are obtained by removing two letters from the $L$-loop word. 

In this paper, motivated by the success of the strike-out experiments \citep{Cai:2024znx}, we investigate whether some of the structures implicitly found by the machine can be discovered analytically and explicitly. Indeed, we find numerous interesting examples of simple relations between coefficients of different words, sequences of words whose coefficients we can determine at \emph{any} loop order, as well as recursion relations that they satisfy.
Together with known linear relations from dihedral symmetry, branch-cut discontinuities, integrability, and final-entry relations \citep{Dixon:2022rse,Cai:2024znx}, 
 such all-loop, recursive results could be extremely useful to assist an AI model in generalizing to all loop orders.
 Alternatively, the simple relations and all-loop sequences could be used to solve the bootstrap equations analytically without an explosion of unknowns in the system of linear equations.

The remainder of this paper is structured as follows.
In Sec.~\ref{sec:formfactor}, we give a brief account of the three-gluon form factor, its symbol, and its known properties.
We describe simple relations among coefficients of different words at the same loop order in Sec.~\ref{sec:empiricalrelations}.
In Sec.~\ref{sec:rays},  we identify sequences of integer coefficients within the three-gluon form factor that can be inferred to all loop orders. 
We provide the full list of all-loop sequences we have found in an ancillary file, \texttt{all8\_rest\_f.txt}. 
In Sec.~\ref{sec:recursion}, we describe the recursion relations we found.
We conclude in Sec.~\ref{sec:conclusions} with an outlook on future directions.
Additional details of our calculations are given in a number of appendices.

\section{The three-gluon form factor and its symbol}\label{sec:formfactor}

In this section, we briefly introduce the form factor we will study throughout this paper, as well as its symbol; see refs.~\citep{Brandhuber:2011tv,Brandhuber:2012vm,Dixon:2020bbt,Dixon:2022rse} for more in-depth background information.

We study the three-gluon form factor of the so-called chiral half of the stress tensor supermultiplet in ${\cal N}=4$ SYM theory. This supermultiplet contains, in particular, the operator $\tr(F_+^2)$, which closely resembles the operator $\tr(F^2)$ 
that couples the Higgs boson to gluons in the limit of a large top quark mass in QCD~\citep{Wilczek:1977zn,Shifman:1978zn}.  Thus, the three-gluon form factor of $\tr(F_+^2)$ is an analog, in planar $\mathcal{N}=4$ SYM theory, of the Higgs-to-three-gluon amplitude in the large-top-mass limit of QCD~\citep{Gehrmann:2011aa,Brandhuber:2012vm}.  Referring to the momenta of the three gluons as $p_i$ with $i=1,2,3$, and to the momentum associated to the operator as $q=p_1+p_2+p_3$, the three-gluon form factor depends on the dimensionless ratios
\begin{equation}
    \label{eq: definition u,v,w}
    u=\frac{(p_1+p_2)^2}{q^2}\,,\qquad v=\frac{(p_2+p_3)^2}{q^2}\,,\qquad w=\frac{(p_3+p_1)^2}{q^2}\,,
\end{equation}
where $u+v+w=1$.
We will always work in the planar limit of a large number of colors $N_c$; we thus consider the leading-$N_c$ contribution to the color-ordered form factor.

Since infrared (IR) divergences are universal and exponentiate, we consider a particular IR-finite part of the form factor, the so-called Bern-Dixon-Smirnov-like remainder function $\mathcal{F}$ \citep{Bern:2005iz}; see refs.~\citep{Brandhuber:2012vm,Dixon:2020bbt,Dixon:2022rse} for its precise definition.  Henceforth, ``form factor'' will always refer to this function.  Its perturbative expansion in the planar coupling $g^2 \equiv N_c g_{\rm YM}^2/(16\pi^2)$ is 
\begin{equation}
    \mathcal{F}(u,v) = 1 + \sum_{L=1}^\infty g^{2L} \, \mathcal{F}^{(L)}(u,v) \,,
\end{equation}
where $L$ is the loop order.

Conjecturally, the form factor $\mathcal{F}^{(L)}$ at any loop order $L$ is a two-dimensional harmonic polylogarithm (2dHPL) \citep{Gehrmann:2000zt} of transcendental weight $2L$. 2dHPLs, a particular subclass of multiple polylogarithms~\citep{Goncharov:1998kja,Goncharov:2001iea},
satisfy intricate identities; see Ref.~\citep{Duhr:2014woa} for a review. Hence, it has proven to be advantageous to represent these functions in terms of their co-products, and in particular via their maximally iterated co-product, the so-called symbol~\citep{Goncharov:2010jf}.
The symbol $\mathcal{S}(F)$ of a multiple polylogarithm $F$ of transcendental weight $n$ is defined recursively via its derivative:
\begin{equation}
    \label{eq: definiton symbol}
    dF=\sum_{i}F_i \ d\log l_i \qquad \Rightarrow \qquad\mathcal{S}(F)=\sum_{i}\mathcal{S}(F_i)\otimes  l_i\,,
\end{equation}
where $F_i$ are multiple polylogarithms of weight $n-1$. The $l_i$ are referred to as symbol letters and $\mathcal{L}=\{ l_i\}_i$ is referred to as the symbol alphabet.
The three-gluon form factor has an alphabet consisting of only six letters, which we label $\{a,b,c,d,e,f\}$:
\begin{equation}
\label{eq: definition abcdef}
    a=\sqrt{\frac{u}{vw}}\,,\quad
    b=\sqrt{\frac{v}{uw}}\,,\quad
    c=\sqrt{\frac{w}{uv}}\,,\quad
    d=\frac{1-u}{u}\,,\quad
    e=\frac{1-v}{v}\,,\quad
    f=\frac{1-w}{w}\,.
\end{equation}
While the symbol does not uniquely specify the full transcendental function,  the ambiguities belong to a considerably smaller space that can be treated similarly; thus, most of the computational work performed in finding the function consists of determining its symbol.

At one-loop order, 
the symbol of the three-gluon form factor is
\begin{equation}
    \label{eq: one-loop example}
    \begin{aligned}
    \mathcal{S}[\cF^{(1)}] = (-2) \Bigl[ b \otimes d + c \otimes e + a \otimes f
  + b \otimes f + c \otimes d + a \otimes e \Bigr] \,,        
    \end{aligned}
\end{equation}
while at two-loop order, 
it is given by
\begin{equation}
    \label{eq: two-loop example}
   \begin{aligned}
       \mathcal{S}[\cF^{(2)}] =&\ {}8 \Bigl[
    b \otimes d \otimes d \otimes d
  + c \otimes e \otimes e \otimes e
  + a \otimes f \otimes f \otimes f \\
&\hphantom{{}8 \Bigl[}
  + b \otimes f \otimes f \otimes f
  + c \otimes d \otimes d \otimes d
  + a \otimes e \otimes e \otimes e \Bigr] \\
&
+ 16 \Bigl[
    b \otimes b \otimes b \otimes d
  + c \otimes c \otimes c \otimes e
  + a \otimes a \otimes a \otimes f \\
&\hphantom{ + 16 \Bigl[}
  + b \otimes b \otimes b \otimes f
  + c \otimes c \otimes c \otimes d
  + a \otimes a \otimes a \otimes e \Bigr]\,.
   \end{aligned}
\end{equation}
More generally, we can write 
\begin{equation}
    \label{eq: symbol in general}
    \mathcal{S}[{\cF}^{(L)}] =
       \sum_{l_{i_1},\ldots,l_{i_{2L}}\in\cL} C^{l_{i_1},\ldots, l_{i_{2L}}} \,
l_{i_1} \otimes \cdots \otimes l_{i_{2L}} \,,
\end{equation}
where the rational coefficients $C^{l_{i_1},\ldots, l_{i_{2L}}}$ are integers in the conventions given by eq.~\eqref{eq: definition abcdef}.\footnote{Integrality of the coefficients is the reason we introduced the letter normalization~\eqref{eq: definition abcdef} in Ref.~\citep{Cai:2024znx}; the earlier normalization of 
Ref.~\citep{Dixon:2022rse} leads to non-integer coefficients.}
Rephrasing, we can consider the symbol at loop $L$ as a sum of $6^{2L}$ \emph{words}, sequences of $2L$ letters drawn from the alphabet $\{a,b,c,d,e,f\}$, multiplied by integer \emph{coefficients}. 
In the remainder of the text, we will frequently drop the tensor product in the words for the sake of brevity, thus we write for example $\tt aaaf$ instead of $ a\otimes a\otimes a\otimes f$.

The symbol at every loop order is left invariant by the dihedral group $D_3 \equiv S_3$, which is generated by two elements,
\begin{equation}
   \begin{aligned}
&\text{\textbf{cycle:} } &&\{{\tt a}, {\tt b}, {\tt c}, {\tt d}, {\tt e}, {\tt f}\} \to \{{\tt b}, {\tt c}, {\tt a}, {\tt e}, {\tt f}, {\tt d}\}\,, \\
&\text{\textbf{flip:} } &&\{{\tt a}, {\tt b}, {\tt c}, {\tt d}, {\tt e}, {\tt f}\} \to \{{\tt b}, {\tt a}, {\tt c}, {\tt e}, {\tt d}, {\tt f}\} \,.
\end{aligned}
\label{dihedralsymm}
\end{equation}
This dihedral symmetry is a consequence of Bose symmetry, acting on the three gluons.

\subsection{Adjacency constraints}

For most of the $6^{2L}$ words at $L$-loop order, the associated coefficients are zero; see table \ref{tab:elem_count1}. For instance, out of $2.8\cdot 10^{12}$ words in the eight-loop symbol, only $1.67\cdot 10^9$ ($0.06\%$) have nonzero coefficients. We refer to the words associated with zero coefficients as \emph{zero elements.} 

Most zero elements can be accounted for by four \emph{adjacency rules}.  Specifically, words with nonzero coefficients must
\begin{enumerate}[label=(\roman*),nosep]
\item begin with $\tt a$, $\tt b$ or $\tt c$, \label{initialrule}
\item end with $\tt d$, $\tt e$ or $\tt f$, \label{finalrule}
\item not have adjacent $\tt a$ and $\tt d$, $\tt b$ and $\tt e$, or $\tt c$ and $\tt f$, \label{firstadjacency} as well as
\item not have adjacent $\tt d$ and $\tt e$, $\tt d$ and $\tt f$, or $\tt e$ and $\tt f$. \label{secondadjacency}
\end{enumerate}
Rule \ref{initialrule} is a branch-cut condition~\citep{Dixon:2022rse}.  The other three rules follow from \emph{antipodal duality} \citep{Dixon:2021tdw} as causal properties of the dual space of hexagon functions describing six-gluon scattering amplitudes~\citep{Dixon:2011pw,Dixon:2013eka,Caron-Huot:2019bsq}.  Antipodal duality is presently an empirical observation.

Table~\ref{tab:elem_count1} presents the number of adjacency-allowed and nonzero elements for all loops up to $8$. At high loops, the adjacency rules account for most of the zero elements, and only about half of the adjacency-allowed words are zero elements. We give a method for calculating these numbers in appendix~\ref{app:combinatorics}.  Other rules, discussed below, will account for most of the remaining zeroes.

\begin{table}[t]
    \small
    \centering
    \begin{tabular}{lccccccccc}
        \toprule
        & $L=1$ & $L=2$ & $L=3$ & $L=4$ & $L=5$ & $L=6$ & $L=7$ & $L=8$  \\
        \midrule
        Total ($6^{2L}$) & $36$ & $1296$ & 46,656 & $1.7\cdot 10^6$ & $6.0\cdot 10^7$ & $2.2 \cdot 10^9$ & $7.8\cdot 10^{10}$ & $2.8\cdot 10^{12}$\\
        Adjacency-allowed & $6$ & $102$ & 1830 & 32,838 & 589,254 & $1.1\cdot 10^7$ & $1.9\cdot 10^8$ & $3.4\cdot 10^9 $\\
        Nonzero & $6$ & $12$ & $636$ & 11,208 & 263,880 & $4.9\cdot 10^6$ & $9.3\cdot 10^7$ & $1.67\cdot 10^9$\\
       \bottomrule
    \end{tabular}
    \caption{\small Breakdown of the total number of symbol elements at loops 1 to 8 into adjacency-allowed elements and the actual nonzero ones.}
    \label{tab:elem_count1}
\end{table}

The four adjacency rules impose strict conditions on the structure of words for nonzero elements (and adjacency-allowed zero elements). Rule \ref{secondadjacency} imposes that the three letters ${\tt d}$, ${\tt e}$ and ${\tt f}$ only appear as \emph{runs} of one or more identical letters; e.g.\ the sub-word ${\tt fff}$ is a run that may appear in a nonzero element, but the sub-word ${\tt efe}$ cannot appear. No such constraint exists on the letters ${\tt a}$, ${\tt b}$ and ${\tt c}$, which can appear in any succession. As a result, adjacency-allowed words decompose into a succession of sequences of the three letters ${\tt a}$, ${\tt b}$ and ${\tt c}$ ($abc$-sequences), interspersed by runs of one of the three letters ${\tt d}$, ${\tt e}$ or ${\tt f}$. Words must start with an $abc$-sequence (rule \ref{initialrule}) and end with a run (rule \ref{finalrule}). Finally the letter in each run is constrained by rule \ref{firstadjacency}: for a run in between two $abc$-sequences ending and beginning with different letters (e.g.\ ${\tt abc\underline{ddd}bca...}$), only one letter is possible. For the last run, or for runs between two $abc$-sequences ending and beginning with the same letter, two letters are possible. 
We can therefore represent any word that satisfies the adjacency relations as a succession of $abc$-sequences and runs: e.g.\ ${\tt abdddccddd}$ as ${\tt ab}$/${\tt d}^3$/${\tt cc}$/${\tt d}^3$. This \emph{run representation} will prove useful when characterizing properties of the symbol.

\subsection{Other homogeneous linear relations}

In addition to the rules \ref{initialrule}--\ref{secondadjacency}, the symbol elements satisfy a number of homogeneous linear equations. Four classes of such homogeneous equations have previously been identified.
The first class of relations involves \emph{pairs} of adjacent letters and takes the form%
\footnote{Note that we are using a different, more word-inspired notation compared to refs.~\citep{Dixon:2022rse,Cai:2024znx}, where the corresponding relations were written as $\sum_{i,j}C_{ij} F^{x_i,x_j}=0$.}
\begin{equation}
\label{eq: integrability schematically}
    \sum_{i,j=1}^6 C_{ij} \, c({\tt Y}{\tt x}_i{\tt x}_j{\tt Z})=0\,,
\end{equation}
where $C_{ij}$ are specific sets of 36 small integers (most of which are zero), ${\tt x}_i,{\tt x}_j$ denote letters in the 6-letter alphabet $\{{\tt a},{\tt b},{\tt c},{\tt d},{\tt e},{\tt f}\}$, {\tt Y} and {\tt Z} are sub-words in these letters of \emph{any} length, and $c({\tt Yx}_i{\tt x}_j{\tt Z})$ denotes the coefficient of the corresponding word. 
All such relations that hold for arbitrary {\tt Y} and {\tt Z} are known~\citep{Dixon:2020bbt,Dixon:2022rse}. They follow from functional integrability, i.e.\ from the fact that partial derivatives commute, and they hold for any choice of sub-words ${\tt Y}$ and ${\tt Z}$ as well as any loop order. For example,\footnote{The other linear relations from functional integrability are given in appendix \ref{app: abc}.}
\begin{equation}
    c({\tt YabZ})+c({\tt YacZ})-c({\tt YbaZ})-c({\tt YcaZ}) =0\,,
\label{eq:intexample}
\end{equation}
which corresponds to setting $C_{12}=C_{13}=1$, $C_{21}=C_{31}=-1$, and the remaining $C_{ij}=0$ in eq.~\eqref{eq: integrability schematically}.

The second class of relations involves adjacent \emph{triplets} of letters, and is of the form\footnote{These relations are given fully explicitly in appendix \ref{app: abc}.} 
\begin{equation}
\label{eq: triple relations schematically}
    \sum_{i_1,i_2,i_3=1}^6 C_{i_1i_2i_3} \, c({\tt Y}{\tt x}_{i_1}{\tt x}_{i_2}{\tt x}_{i_3}{\tt Z})=0\,,
\end{equation}
where the analogous $C_{i_1i_2i_3}$ are again specific sets of small integers (mostly zero), and the notation is as before.
Again, all such relations of this class that hold for any choice of sub-words ${\tt Y}$ and ${\tt Z}$ as well as any loop order are known \citep{Dixon:2020bbt,Dixon:2022rse}. They follow from functional integrability for the six-gluon amplitude, when combined with the conjectural antipodal duality \citep{Dixon:2021tdw}. We refer to eqs.~\eqref{eq: integrability schematically} and \eqref{eq: triple relations schematically} as {\it generic pair} and {\it triple} relations, because they hold at arbitrary positions within the symbol, because both {\tt Y} and {\tt Z} are of arbitrary length.

The third and fourth classes of homogeneous relations are of the form 
\begin{equation}
\label{eq: initial}
    \sum_{i_1,\dots,i_k=1}^6 \hat{C}_{i_1\dots i_k} \, c({\tt x}_{i_1}{\tt\dots x}_{i_k}{\tt Z})=0
\end{equation}
and 
\begin{equation}
\label{eq: final}
    \sum_{i_1,\dots,i_k=1}^6 \tilde{C}_{i_1\dots i_k} \, c({\tt Y}{\tt x}_{i_1}{\tt \dots x}_{i_k})=0\,.
\end{equation}
We refer to eq.~\eqref{eq: initial} as (multiple) {\it initial}-entry conditions, because they involve specific combinations of strings of $k$ letters at the beginning of the symbol, followed by anything ({\tt Z}) at the end.
Most of the initial-entry conditions \eqref{eq: initial} follow from the (single) initial-entry (branch-cut) condition \ref{initialrule}, the pair relations \eqref{eq: integrability schematically}, and the triple relations \eqref{eq: triple relations schematically}, as well as further branch-cut conditions described in appendix \ref{app: bootstrapping}. The solutions to these relations have been worked out up to $k=8$~\citep{Dixon:2022rse}. Moreover, they contain empirical relations starting at $k=4$ that were observed in the data up to $L=8$.
The multiple {\it final}-entry conditions \eqref{eq: final} follow from the multiple-initial-entry relations and functional integrability relations for the six-gluon amplitude, via antipodal duality~\citep{Dixon:2021tdw}. Again they are labelled by an integer $k$, which now specifies how many letters at the \emph{end} of the symbol are involved. They can in principle be worked out to any $k$, and this has been done explicitly up to $k=11$ by using antipodal duality~\citep{Caron-Huot:2019bsq}.

\section{Simple linear relations}\label{sec:empiricalrelations}

In the following, we will discuss several simple linear relations, which we observed empirically. In contrast to the previously mentioned relations discussed above, they involve sub-words of arbitrary length.

\subsection{Zero-suffix rules}
\label{sec:zerosuffix}

The first group of such relations, which we initially observed empirically, reads as follows:
\begin{equation}
\label{eq:generalized zero suffices}
    \begin{aligned}
        c({\tt X ba\dots af})=0\,,\\
        c({\tt X ca\dots af})=0\,,\\    
    \end{aligned}
\end{equation}
as well as its images under dihedral symmetry. Here, ${\tt a\dots a}$ denote \emph{one or more} repetitions of the letter ${\tt a}$, and ${\tt X}$ is an arbitrary sub-word. These relations impose that a nonzero word ending with a run of one letter (e.g.\ with ${\tt af}$) must have only one letter in its last $abc$-sequence (i.e.\ it can end with ${\tt aaaf}$ but not ${\tt baaf}$, even though adjacency relations allow it). These \emph{zero-suffix} relations follow from the relations described in Sec.~\ref{sec:formfactor}; see appendix \ref{app:gzsproof} for a proof.
They are specific to words ending with a run of length one. No similar rules exist for words with final runs of length $2$ to $8$.  

The rules \eqref{eq:generalized zero suffices} account for all zero-suffixes observed in prior work~\citep{Dixon:2022rse} and of the form \eqref{eq: final}, and more.  As shown in appendix~\ref{app:combinatorics}, table~\ref{tab:k_allowed}, the terms allowed by these rules are fairly exhaustive: by loops 7 and 8, over $92\%$ of the terms allowed to be nonzero are actually nonzero in the symbol.  We hypothesize that the only coefficient zeroes that can be associated with final strings are precisely eqs.~\eqref{eq:generalized zero suffices} and no more. We call this the \emph{one-run zeroes hypothesis}. In appendix~\ref{app:combinatorics}, we count the number of allowed terms under this hypothesis.   We show that there are auxiliary sequences associated with the combinatorics which obey Fibonacci-like recursion relations with an asymptotic growth rate of $2+\sqrt{5}$ per weight, or $(2+\sqrt{5})^2 \approx 17.94$ per loop.  Because the number of actual nonzero symbol terms appears to approach a fixed fraction of the allowed terms, we conjecture that the asymptotic growth rate of the number of terms in the symbol is also $(2+\sqrt{5})^2$ per loop.

\subsection{Zero-prefix rules}
\label{sec:zeroprefix}

Having discussed the implications of zero-suffix rules, we now turn to \emph{zero-prefix} rules.  The only zero-prefix rules we have found are
\begin{equation}
\label{eq: prefix zero}
    c({\tt abd Z})=0,
\end{equation}
as well as its dihedral images, for any sub-word ${\tt Z}$. These rules involve an initial $abc$-sequence of length $2$, followed by a run for a letter that is incompatible with the first letter of the word (i.e~${\tt a}$ and ${\tt d}$, ${\tt b}$ and ${\tt e}$, or ${\tt c}$ and ${\tt f}$). Relation \eqref{eq: prefix zero} is a special case of the known relations \eqref{eq: initial}, in which only one of the elements of the set $\hat{C}_{i_1\ldots i_k}$ is nonzero.
Table~\ref{tab:elem_count} presents a breakdown of the zero and nonzero elements for loops $1$ to $8$; see appendix~\ref{app:combinatorics} for a method for calculating these numbers.
The prefix and suffix rules account for more than $99.7\%$ of the zeroes among adjacency-allowed elements above 6 loops. 
That is, there are very few zeroes in the high-loop order symbol that are not explained by the adjacency, prefix or suffix rules.

The zero-prefix and zero-suffix rules concern words with a short (one letter) last run or a short (2 letter) first $abc$-sequence.  They are not ``new'', in the sense that they can be inferred from previously known rules. However, some of them (namely, the suffix rules~\eqref{eq:generalized zero suffices} at large weight) have not been imposed in the traditional bootstrap approach, so they might lead to additional speed-up.

\begin{table}[t]
    \small
    \centering
    \begin{tabular}{lccccccccc}
        \toprule
        & $L=1$ & $L=2$ & $L=3$ & $L=4$ & $L=5$ & $L=6$ & $L=7$ & $L=8$  \\
        \midrule
        Adjacency-allowed & $6$ & $102$ & 1830 & 32,838 & 589,254 & $1.1\cdot 10^7$ & $1.9\cdot 10^8$ & $3.4\cdot 10^9 $\\
        Suffix-forbidden & $0$ & $48$ & $864$ & 15,504 & 278,208 & $5.0\cdot 10^6$ & $9.0 \cdot 10^7$ & $1.6 \cdot 10^{9}$ \\
        Prefix-forbidden & $0$ & $6$ & $78$ & 1,206 & 21,438 & $3.8\cdot 10^5$ & $6.9\cdot 10^6$ & $1.2\cdot 10^8$ \\
        \midrule
        Remaining allowed & $6$ & $48$ & $888$ & 16,128 & 289,608 & $5.2\cdot 10^6$ & $9.3\cdot 10^7$ & $1.67\cdot 10^9$  \\
        Zeroes & $0$ & $36$ & $252$ & 4,920 & 25,728 & $2.8 \cdot 10^5$ & $3.0 \cdot 10^5$ & $1.77 \cdot 10^6$  \\
        Nonzeroes & $6$ & $12$ & $636$ & 11,208 & 263,880 & $4.9\cdot 10^6$ & $9.3\cdot 10^7$ & $1.67\cdot 10^9$\\
       \bottomrule
       Nonzero fraction & 1 & 0.25 & 0.7162 & 0.6949 & 0.9112 & 0.9460 & 0.9968 & 0.9989 \\
       \bottomrule
    \end{tabular}
    \caption{\small 
    Number of adjacency-allowed elements at loops 1 to 8 of table \ref{tab:elem_count1}, followed by those forbidden by the suffix rule \eqref{eq:generalized zero suffices}. Of those remaining, the number forbidden by the prefix rule \eqref{eq: prefix zero} is given next. The remaining allowed elements are further broken down to zero elements and nonzero elements in the actual $L$-loop symbol.  The fraction of remaining allowed elements that are actually nonzero is over 99.6\% by 7 loops!
    }
    \label{tab:elem_count}
\end{table}

\subsection{Two-term relations}
\label{sec:twoterm}

A second group of empirical final-entry relations holds between coefficients of two different sets of terms.  For all $n>1$ and $1<k<n$, we find that
\begin{equation}
\label{eq: powerful linear relation}
    c({\tt Xfaf}^{n-1})=c({\tt Xf}^k{\tt af}^{n-k})\,,
\end{equation}
as well as its images under dihedral symmetry.
We have observed that this rule holds for all symbols up to eight-loop order, and we conjecture it to be true in general.
We checked that the relations \eqref{eq: powerful linear relation} for $n\leq10$ are implied by the multiple-final-entry relations \eqref{eq: final}, in the sense that they are automatically satisfied once the multiple-final relations have been imposed.\footnote{
For $n=8,9,10$ we used antipodal duality and the hexagon function space to perform this check.}
Thus, it should be possible to prove eq.~\eqref{eq: powerful linear relation} via the initial-entry condition and functional integrability for the antipodally-related hexagon function space describing the six-gluon amplitude; we leave this for future work. Again, while they do not give genuinely new constraints, the relations \eqref{eq: powerful linear relation} could be used for a more efficient solution than the traditional bootstrap.

A third group of simple relations holds for any sub-words $\tt Y$ and $\tt Z$ and any $abc$-sequence $\tt A$, i.e.\ a sub-word $\tt A$ only including the letters $\tt a,b,c$:
\begin{equation}
c({\tt YfaAbfZ}) = c({\tt YfbAafZ}) = -c({\tt YfbAbfZ})\,,
\label{eq:aAbswap}
\end{equation}
as well as its dihedral images. We prove the relations~\eqref{eq:aAbswap} in appendix~\ref{app: abc}.
Again, while eq.~\eqref{eq:aAbswap} is not a genuinely new equation, it is particularly simple and so it may lead to more insights, for example in the context of relating different all-loop sequences, of the type discussed in the next section.

\section{All-loop sequences}
\label{sec:rays}

In this section, we identify numerous sequences of terms in the symbol that we can write in a closed form at any loop order.

For example, the symbol of the three-gluon form factor contains at each loop order $L$ a term ${\tt a\ldots af}$, where the letter ${\tt a}$ is repeated a total of $2L-1$ times in a row to result in a total of $2L$ letters.
The coefficient of this term, which we denote by $c_L$,  is the largest coefficient at loops 1 to 8, and presumably at any loop order.%
\footnote{Heuristically, it is not surprising that the largest coefficient is associated to a word with the largest repetition of a letter, because the symbol gives a factorial enhancement to repeated letters. For example, the symbol of $\log(x)^n$ is the word $\tt x\dots x$ with coefficient $n!$.} It is found to be 
\begin{equation}
c_{L=1,2,3,4,5,6,\dots}({\tt a\ldots af}) = -2, 16, -384, 15360, -860160, 61931520, \ldots\,.
\end{equation}
Up to the alternating sign, this sequence is entry \texttt{A052737} in the online encyclopedia of integer sequences \url{oeis.org}~\citep{OEISA052737}, which yields the simple all-loop formula
\begin{equation}
c_L({\tt a\ldots af}) = 
\frac{(-4)^L}{2} \frac{[2(L-1)]!}{(L-1)!}=
(-1)^L \, 2^{3L-2} \, (2L-3)!! \,,
\label{eq:Onebraycoeff}
\end{equation}
where $n!!=\prod_{k=0}^{\lceil n/2\rceil-1}(n-2k)$ denotes the double factorial of $n$.  This formula is easily confirmed using the results at 7 and 8 loops.
Similarly, we find  
\begin{equation}
c_L({\tt af\ldots f}) = 
(-1)^L \, 2^{2L-1} \, (2L-3)!! \,.
\label{eq:bdddd}
\end{equation}

Focusing on more general sequences that begin with $m$ arbitrary letters and end with $2L-m$ letters $\tt{f}$, we observed the following interesting pattern. Whenever we could determine them, they were given by a factor of
\begin{equation}
    (-1)^L\, 2^{2L-2\lceil{m/2}\rceil} \, (2L-1-2\lceil{m/2}\rceil)!!
\end{equation}
multiplied by a polynomial in $L$ of degree $\lceil{m/2}\rceil-1$; i.e.\ degree $0,0,1,1,2,2,3,3,\dots $ for $m=1,2,3,4,5,6,7,8,\ldots$. 
The shortest element of a sequence is the first one with at least one $\tt f$. We are guaranteed to encounter the shortest element of a sequence provided that $m+1\leq2L$. For $m=1,2,3,4,5,6,7,8$ that will happen by loop order $1,2,2,3,3,4,4,5$, i.e.~by loop order $\lfloor{m/2}\rfloor+1$. Therefore, if we know the first $L$ loop orders, we can use loop orders from $\lfloor{m/2}\rfloor+1$ up to $L$ to determine the polynomial.  That constitutes $L-\lfloor{m/2}\rfloor$ data points, and we want to fix a polynomial in $L$ of degree $\lceil{m/2}\rceil-1$, which requires $\lceil{m/2}\rceil$ data points. We obtain a unique answer if $L-\lfloor{m/2}\rfloor\geq \lceil{m/2}\rceil$, i.e.\ if $L\geq m$. Moreover, we have at least one cross check if $L>m$.
In particular, with the full $L\leq8$ data set available to us, we could uniquely determine all coefficients in the ansatz until $m=8$, with at least one cross check for $m\leq 7$.

Using this approach, we have determined all-loop expressions for any sequence terminated by $2L-8$ {\tt f}'s,
\begin{equation}
    c_L({\tt X}_8 {\tt f\ldots f}) 
   = p_L({\tt X}_8  {\tt f\ldots f}) \times (-1)^L 2^{2 L - 8} (2 L - 9)!!\,,
\end{equation}
where $p_L({\tt X}_8 {\tt f\ldots f})$ is a polynomial in $L$ and, for simplicity, we have chosen one common factor for all sequences.
Here are some examples of the corresponding polynomials:
\begin{align}
p_L({\tt aaf\ldots f}) &= 0\,,  \label{eq:aafXf}\\
p_L({\tt caaf\ldots f}) &= 32(L-2)(2L-5)(2L-7) \,, \label{eq:caafXf}\\
p_L({\tt caaaf\ldots f}) &= \frac{16}{3}(4L-9)(2L-5)(2L-7) \,, \label{eq:caaafXf}\\
p_L({\tt ccaaaaf\ldots f}) &= -\frac{4}{5}(2L-7)(7L^2+22L-140) \,,\\
p_L({\tt cccccaaf\ldots f}) &= -\frac{8}{3}(L-4)(L^2-47L+135) \,,\\
p_L({\tt aeeeaaaf\ldots f}) &= -\frac{2}{45}(163 L^3 - 2220 L^2 + 15977 L - 36660) \,.
\end{align}
These equations are understood to hold for all loop orders $L$ for which there is at least one $\tt f$ present in the final run.
Note that for $L\leq4$, appropriate factors in the polynomials should be absorbed in the double factorials in order to avoid double factorials of negative arguments.%
\footnote{The polynomials $p_L$ can contain non-integer rational-number coefficients, and can even evaluate to non-integers.  However, the full sequence, including the common factor, $(-1)^L 2^{2 L - 8} (2 L - 9)!!$, is guaranteed to be integer-valued for integers $L\geq5$.}

We provide the full set of sequences in the ancillary file \texttt{all8\_rest\_f.txt}.  
There are 42,376 nonzero sequences in this file.  However, far fewer of them are actually independent.  There are only 1,251 linearly independent weight-8 symbols allowed by the multi-initial-entry conditions.  We require that these symbols do not end in 
{\tt c}, {\tt d}, or {\tt e} -- due to the adjacency condition with {\tt f} in the $9^{\rm th}$ slot.  We also impose invariance under the {\tt f}-preserving flip in eq.~\eqref{dihedralsymm}. Then we are left with only 261 independent combinations.

As we will discuss in the conclusion and elaborate on in appendix \ref{app: bootstrapping}, these all-loop sequences can be used successfully to replace almost all inhomogeneous constraints from the FFOPE.
If we are at loop order $L$, then we can use all the information from loop order $L-1$.  According to the discussion above, at loop order $L-1$ we can determine all sequences with $m=L-1$ arbitrary letters, followed by $(L-1)$ {\tt f}'s.  At the next loop order, $L$, we add two more {\tt f}'s to the back of the sequence, so we should know all sequences ending in $(L+1)$ {\tt f}'s.

\section{Recursion relations} \label{sec:recursion}

In the strike-out experiments in our machine-learning study \citep{Cai:2024znx}, we found evidence that much information about a coefficient $c_L({\tt\dots})$ can be recovered from the list of coefficients of its \emph{strike-two parents}, which are obtained by deleting two letters of the word to arrive at a word at $L-1$ loops. Defining the distance between the two struck-out letters to be $k$, we found that reducing $k$ as far as to $k=2$ does not lead to a significant drop in the accuracy of the predictions, which was above $98\%$.
Motivated by these findings, in this section we strive to find \textit{exact} and \textit{analytic} relations between coefficients of words and the coefficients of their strike-two parents at one lower loop. 
To this end, we use the all-loop expressions found in the previous section, in order to fit coefficients in a corresponding ansatz.

We start with the sequence ${\tt af\dots f}$, which is special because its only non-vanishing parents are obtained by striking out two ${\tt f}$s, resulting in ${\tt af\dots f}$ at one loop order less.
Indeed, we find from eq.~\eqref{eq:bdddd} that
\begin{equation}
    c_L({\tt af\dots f}) = (12 - 8 L) c_{L-1}({\tt af\dots f})\,.
    \label{eq:recuraf}
\end{equation}
The ``$-8L$'' part appears to be universal for sequences ending in ${\tt f}$s, in the sense that it is present in all relations we found, see below.

Note that the number of ways to strike out two different ${\tt f}$s separated by a fixed, finite distance $k$ grows linearly in $L$. (Whereas the number of ways to strike out an {\tt f}, and also some other letter appearing before the final {\tt f}, will not grow with $L$ for finite $k$.)  If we assume that there are indeed general recursive formulas with fixed but finite $k$, then the linear-in-$L$ factor in eq.~\eqref{eq:recuraf} can be explained heuristically.

We also find from eq.~\eqref{eq:Onebraycoeff} that
\begin{equation}
    c_L({\tt a\dots a f}) = (24 - 16 L) c_{L-1}({\tt a\dots af})\,.
\end{equation}
The linear-in-$L$ growth here could be explained similarly by counting the number of ways of striking out two different ${\tt a}$s. 

The word  ${\tt abf\dots f}$ has a vanishing coefficient for all $L$.  This vanishing is not a consequence of a zero-suffix, zero-prefix, or any other linear relation from Secs.~\ref{sec:formfactor} and \ref{sec:empiricalrelations}.  On the other hand, ${\tt abf\dots f}$ has two strike-out parents with non-vanishing coefficients, i.e.~when we remove {\tt a} and {\tt f} or remove {\tt b} and {\tt f}.  If we demand that ${\tt abf\dots f}$ participates in a recursion relation and fix coefficients in a corresponding ansatz, we find 
\begin{equation}
    0=c_L({\tt abf\dots f})= C \times [ c_{L-1}({\tt af\dots f})- c_{L-1}({\tt bf\dots f})]\,,
  \label{eq:recurabf}  
\end{equation}
where the constant $C$ remains undetermined since $c_{L-1}({\tt af\dots})= c_{L-1}({\tt bf\dots})$ due to dihedral symmetry.
Below, we will encounter further undetermined constants due to dihedral symmetry or other known linear relations.

Next, we inspect $abc$-sequences of length three, followed by {\tt f}s.
These two sequences have only two nonzero parents, whose coefficients in a recursion relation are uniquely determined:
\begin{align}
    c_{L}({\tt caaf\dots f}) &= (20 - 8 L) c_{L-1}({\tt caaf\dots f}) - c_{L-1}({\tt af\dots f})\,,\\
    c_L({\tt cdbf\dots f}) &= (20 - 8 L) c_{L-1}({\tt cdbf\dots f}) + 2 c_{L-1}({\tt bf\dots f})\,.
\end{align}
There is no linear-in-$L$ term in the second summand on the right-hand side, which is in line with the number of ways of performing a finite $k$ strike-two procedure.
In the following case,
\begin{equation}
    c_{L}({\tt cabf\dots f}) = (20 - 8 L) c_{L-1}({\tt cabf\dots f}) + (1 - C) c_{L-1}({\tt 
af\dots f}) + C c_{L-1}({\tt bf\dots f})\,,
\end{equation}
the constant $C$ is again undetermined and drops out due to dihedral symmetry, as in eq.~\eqref{eq:recurabf}.

Proceeding to more non-${\tt f}$-letters, we find
\begin{align}
c_L({\tt ccabf\dots f}) &= (12 - 8 L) c_{L-1}({\tt ccabf\dots f})\,,\\
c_L({\tt cacaf\dots f}) &= (28 - 8 L) c_{L-1}({\tt cacaf\dots f}) + 16 c_{L-1}({\tt caaf\dots f})\,,\\
c_L({\tt cabbf\dots f}) &= (24 - 8 L) c_{L-1}({\tt cabbf\dots f}) - 4 c_{L-1}({\tt cabf\dots f})\,,
\end{align}
showing similar structures as before.
In general, more undetermined coefficients can occur, such as
\begin{equation}
\begin{aligned}
c_L({\tt cccacaf\dots f}) = {}&(7 C - 8 L) c_{L-1}({\tt cccacaf\dots f}) +    (-60 + 15 C) c_{L-1}({\tt cccaaf\dots f}) \\&+ (-102 + 20 C) c_{L-1}({\tt ccaaf\dots f}) + (30 - 6 C) c_{L-1}({\tt cacaf\dots f})\,.
\end{aligned}
\end{equation}

The next interesting observation happens for 
\begin{multline}
    c_L({\tt cccaaf\dots f}) = (28 - 8 L) c_{L-1}({\tt cccaaf\dots f}) + 40 c_{L-1}({\tt ccaaf\dots f}) \\+ 12 c_{L-1}({\tt caaf\dots f})\,,
\end{multline}
where the second term on the right-hand side is a $k=3$ parent but not a $k=2$ parent. An analogous recursion relation without this term does not exist, indicating that $k>2$ is necessary in general.
The need for $k>2$ is consistent with the imperfect accuracy of the ML experiments.

However, the strike-two recursion breaks down altogether for ${\tt cdcaf\dots f}$, which satisfies
\begin{equation}
 c_L({\tt cdcaf\dots f}) =   \left( 12 - \frac{12}{-4 + L} - 8 L\right) c_{L-1}({\tt cdcaf\dots f})\,.
\end{equation}
This recursion has a singularity at $L=4$, which is consistent with the fact that ${\tt cdcaffff}$ at $L=4$ has no strike-two parents with non-vanishing coefficients, i.e.~$c_3({\tt cdcaff}) = c_3({\tt ccafff}) = \ldots = 0$.
However, if we also allow for strike-four parents, corresponding to recursion relations that span more than one loop order, we find a solution,
\begin{equation}
    c_L({\tt cdcaf\dots f}) = (12 - 8 L) c_{L-1}({\tt cdcaf\dots f})   - 4 c_{L-2}({\tt af\dots f})\,.
\end{equation}

It would be interesting to extend the study of these recursion relations to more general sequences, as well as beyond sequences, to find recursion relations that describe all data up to eight loops, as well as to predict the prefactors of the terms in the recursion relations.
We leave these investigations for the future.

\section{Conclusions and future work} \label{sec:conclusions}


In this paper, we have identified several interesting patterns in the symbol of the three-gluon form factor.  We began in Sec.~\ref{sec:formfactor} (and appendix~\ref{app:combinatorics}) by tabulating how many words in the symbol can have nonzero coefficients, using the known adjacency, zero-prefix and zero-suffix conditions. This number is very close to the actual number of nonzero terms in the symbol by 7 loops, and it generically obeys a Fibonacci-like relation, leading to an asymptotic growth rate per loop of $(2+\sqrt{5})^2$.  

We then turned our attention to relations among the nonzero integer coefficients. In Sec.~\ref{sec:empiricalrelations}, we found a number of different types of simple linear relations between the coefficients of different words at the same loop order. Some of these relations could be derived from already known relations, while some are new.  In Sec.~\ref{sec:rays}, we observed that the coefficients of many sequences of words are given by closed-form expressions which we expect to be valid at any loop order $L$. In particular, sequences ending in $2L-m$ repeated letters {\tt f} are given by the double factorial $(2L-1-2\lceil{m/2}\rceil)!!$ multiplied by a polynomial in $L$ of degree $\lceil{m/2}\rceil-1$.  In Sec.~\ref{sec:recursion}, we used the all-loop sequences to identify recursion relations that relate the coefficient of a word at a given loop order to the coefficients of particular words at lower loop orders.  Taken together, these findings indicate that the three-gluon form factor exhibits much richer structure than previously anticipated!  

 The all-loop sequences and recursion relations do not follow in any obvious way from known properties of the form factor.  It would be interesting to be able to derive them from first principles, as well as to derive further classes of identities. In the absence of such a derivation, it would be desirable to develop an automated method, potentially based on machine-learning (e.g.~symbolic regression), to discover further all-loop sequences and recursion relations. In the best case, it might be possible to determine recursion relations and/or sequences that fully determine the three-gluon form factor at any loop order. Even if this is not possible, the existence of the all-loop sequences suggests a new way of solving the bootstrap equations, namely by replacing the inhomogeneous equations from the FFOPE by the inhomogeneous equations from the sequences.

As we discuss in appendix \ref{app: bootstrapping}, it is indeed possible to determine all but one of the unknowns in the ansatz for the symbol at $L$ loops from the known homogeneous linear relations, an inhomogeneous constraint on the $L^{\rm th}$ discontinuity, and the all-loop sequences accessible from the data up to $L-1$ loops.%
\footnote{We have explicitly checked this up to seven-loop order, and we conjecture it to be true in general.}
Moreover, it might be possible to avoid the proliferation of unknowns in the bootstrap by starting with the all-loop sequences and only imposing the known homogeneous constraints afterwards.
For this purpose, the simple linear relations are particularly useful. While they are (conjecturally) implied by previously known relations, they suggest a simpler strategy for solving the homogeneous equations without as much proliferation of unknowns.  The all-loop sequences also constitute simple inhomogeneous ``seed'' data, together with the simple linear relations and other homogeneous equations, that may prove useful for a numerical machine-learning model to predict the next loop order.  We leave these exercises for future work.

Finally, similar simple linear relations, all-loop sequences and recursion relations might also exist for the three-gluon form factor of $\phi^3$ \citep{Basso:2024hlx,Henn:2024pki} as well as the NMHV six-gluon amplitude \citep{Caron-Huot:2019vjl} on the parity-preserving surface, which both share the same alphabet and for which high-loop data is equally available.

\paragraph{Acknowledgements:}
LD thanks James Drummond, \"{O}mer G\"urdo\u{g}an, Andrew McLeod and Karen Yeats for earlier collaboration on recurrent sequences.  
This work was supported in part by the U.S. Department of Energy (DOE) under Award No.~DE-FOA-0002705, KA/OR55/22 (AIHEP). 
LD and TC are additionally supported by the U.S. Department of Energy Award No.~DE-AC02-76SF00515. MW was supported by the research grant 00025445 from Villum Fonden as well as by the Sapere Aude: DFF-Starting Grant 4251-00029B. 
LD is grateful to ETH Z\"urich, the University of Z\"urich, Humboldt University, the Simons Center for Geometry and Physics, Stony Brook University, and the Harvard CMSA for hospitality when some of this work was completed.
LD and MW thank the Munich Institute for Astro-, Particle and BioPhysics (MIAPbP) for hospitality, which is funded by the Deutsche Forschungsgemeinschaft (DFG, German Research Foundation) under Germany's Excellence Strategy – EXC-2094 – 390783311.
Moreover, FC, LD and MW thank the Data Science Institute at the University of Wisconsin-Madison, for hospitality during the final stage of this work.

\appendix

\section{Adjacencies in \texorpdfstring{$abc$}{abc}-sequences}
\label{app: abc}

In this appendix, we describe the generic pair and triple relations, which were given schematically in eqs.~\eqref{eq: integrability schematically} and \eqref{eq: triple relations schematically}. We also show how to combine some of these relations in order to move around the composite letter ({\tt abc}).  The latter relations are then used to prove relation~\eqref{eq:aAbswap}.

Underlying the investigations in this paper are the functional integrability relations of the form \eqref{eq: integrability schematically}, which read \citep{Dixon:2022rse}
\begin{equation}
c({\tt Y ab Z})+c({\tt Y ac Z}) = c({\tt Y ba Z})+c({\tt Y caZ})\,,
\label{commute_left}
\end{equation}
\begin{equation}
c({\tt YcaZ})+c({\tt YcbZ}) = c({\tt YacZ})+c({\tt YbcZ})\,,
\label{commute_right}
\end{equation}
and 
\begin{equation}
\begin{split}
c({\tt YdbZ})+c({\tt YcdZ})+c({\tt YecZ})+c({\tt YaeZ})+c({\tt YfaZ})+c({\tt YbfZ})+2c({\tt YcbZ}) \\
=c({\tt YbdZ})+c({\tt YdcZ})+c({\tt YceZ})+c({\tt YeaZ})+c({\tt YafZ})+c({\tt YfbZ})+2c({\tt YbcZ})\,,
\end{split}
\label{eq:14terms}
\end{equation}
as well as the triple relations of the form \eqref{eq: triple relations schematically}, which read \citep{Dixon:2022rse}
\begin{equation}c({\tt YaabZ})+c({\tt YabbZ})+c({\tt YacbZ})=0\,,
    \label{triplet_rel}
\end{equation}
including all images of this relation under the dihedral operations~\eqref{dihedralsymm}.
Here {\tt Y} and {\tt Z} denote arbitrary sub-words.

Introducing the notation,
\begin{equation}
c({\tt Y(abc) Z})\equiv c({\tt YaZ})+c({\tt YbZ})+c({\tt YcZ})\,,
\end{equation}
the relations~\eqref{commute_left} and~\eqref{commute_right} are equivalent to \begin{equation}
c({\tt Y(abc)aZ})=c({\tt Ya(abc)Z})\,.\label{commute}
\end{equation}
So the composite letter ({\tt abc}) can be commuted past any of the letters {\tt a}, {\tt b} or {\tt c}.  If $\tt A$ is a sequence of the letters $\tt a,b,c$, then for any $\tt A_1$ and $\tt A_2$, such that $\tt A_1A_2=A$, we may commute ({\tt abc}) past either string, obtaining
\begin{equation}
    c({\tt YA_1(abc)A_2Z}) = c({\tt Y(abc)AZ}) = c({\tt YA(abc)Z})\,.
    \label{eq:equalcoeffs}
\end{equation}
In other words, for any partition of a given $abc$-sequence {\tt A} into ${\tt A_1A_2}$, all coefficients $c({\tt YA_1(abc)A_2Z})$ have exactly the same value.

Now from the final-entry condition~\ref{finalrule}, {\tt Z} must contain at least one of {\tt d}, {\tt e}, or {\tt f}, and by adjusting ${\tt A_2}$, one of these letters can be put at the start of {\tt Z}. Assume, without loss of generality, that $\tt Z$ begins with {\tt f}.  Then {\tt cZ} violates the adjacency condition~\ref{secondadjacency} and so we have 
\begin{equation}
    c({\tt YA_1(abc)A_2Z})=c({\tt YA(abc)Z})= c({\tt YAaZ})+c({\tt YAbZ})\,.
\end{equation}

Also, relation~\eqref{triplet_rel} asserts that
\begin{equation}
c({\tt Ya(abc)bZ})=0\,. \label{base_triplet}
\end{equation}
If $\tt A$ includes at least two different letters, then relation~\eqref{base_triplet} applies in at least one position of $\tt A$, and the collections of equal coefficients in \eqref{eq:equalcoeffs} actually all vanish, 
\begin{equation}
    c({\tt YA_1(abc)A_2Z}) = c({\tt Y(abc)AZ}) = c({\tt YA(abc)Z}) = 0\,.
\end{equation}

Besides providing the pair and triple relations, in this appendix we have established many relations between coefficients involving the composite letter ({\tt abc}).

We will now show how the relations~\eqref{eq:aAbswap} given in 
Sec.~\ref{sec:twoterm} follow from the above relations. We repeat eq.~\eqref{eq:aAbswap} for the convenience of the reader:
\begin{equation}
c({\tt YfaAbfZ}) = c({\tt YfbAafZ}) = -c({\tt YfbAbfZ})\,.
\label{eq:aAbswap-app}
\end{equation}

First consider the case that {\tt A} contains only a single letter, and that it is {\tt a}.  (If the single letter is {\tt b}, it is the same by dihedral symmetry.)
Specifically, we let ${\tt A=a}^n$, for $n\geq 1$, and we wish to show that
\begin{equation}
c({\tt Yfa}^{n+1}{\tt bfZ})=c({\tt Yfba}^{n+1}{\tt fZ})=-c({\tt Yfba}^{n}{\tt bfZ}) \,.
\label{eq:anbswap}
\end{equation}
We can derive the first eq.~\eqref{eq:anbswap} from {\tt cf} non-adjacency, plus the commutativity property~\eqref{eq:equalcoeffs} of the composite letter ({\tt abc}):
\begin{equation}
\begin{split}
 c({\tt Yfa}^{n+1}{\tt bfZ}) &= c({\tt Yfa}^{n+1}{\tt (abc)fZ}) - c({\tt Yfa}^{n+2}{\tt fZ}) \\
 &= c({\tt Yf(abc)a}^{n+1}{\tt fZ}) - c({\tt Yfa}^{n+2}{\tt fZ}) = c({\tt Yfba}^{n+1}{\tt fZ}).
\end{split}
\end{equation}
For the second eq.~\eqref{eq:anbswap} we consider the difference of the middle and right sides of the equation, which is $c({\tt Yfba}^n{\tt (abc)fZ})$. We commute ({\tt abc}) to the left until it sits between {\tt b} and {\tt a}. Then we use a dihedral image of eq.~\eqref{base_triplet} to conclude that $c({\tt Yfba}^n{\tt (abc)fZ})$ vanishes.

Next consider the case that $\tt A$ has two different letters.  We first show that
\begin{equation}
c({\tt YfAafZ})+c({\tt YfAbfZ})= 0 \,,
\label{eq:absum1}
\end{equation}
and 
\begin{equation}
c({\tt YfaAfZ})+c({\tt YfbAfZ})= 0 \,.
\label{eq:absum2}
\end{equation}
These results again follow from commuting ({\tt abc}) until it sits between two different letters of type {\tt a,b,c}, and then using eq.~\eqref{base_triplet}.

Combining eqs.~\eqref{eq:absum1} and \eqref{eq:absum2}, we always have
\begin{equation}
c({\tt YfaAbfZ})=c({\tt YfbAafZ}) = -c({\tt YfbAbfZ}) = - c({\tt YfaAafZ}) \,,
\label{eq:aAbswapALT}
\end{equation}
so long as $\tt aAa$ and $\tt bAb $ include two different letters.  Eq.~\eqref{eq:aAbswapALT} includes the case that ${\tt A} = {\tt c}^n$.   
Thus we have established eq.~\eqref{eq:aAbswap} in the general case.

\section{A proof of zero-suffix relations}\label{app:gzsproof}

In this appendix, we prove the relations \eqref{eq:generalized zero suffices}
presented in Sec.~\ref{sec:zerosuffix}, which we repeat here for the convenience of the reader: 
\begin{equation}
\label{eq:generalized zero suffices appendix}
    \begin{aligned}
        c({\tt X ba\dots af})=0\,,\\
        c({\tt X ca\dots af})=0\,.\\    
    \end{aligned}
\end{equation}
They and their dihedral images hold for any sub-word ${\tt X}$.

The validity of rule \eqref{eq:generalized zero suffices appendix} for all length of suffixes and all loops can be proven by induction on the length of the suffixes. Specifically, we will show that, for any $n\geq 1$, words ending in ${\tt ba}^n{\tt f}$ and ${\tt ca}^n{\tt f}$ (and their dihedral equivalents) have zero coefficients. For $n=1$, i.e.\ for suffixes ${\tt baf}$, ${\tt caf}$, this rule is the antipodal dual of allowed-weight three hexagon functions for the six-gluon scattering~\citep{Dixon:2021tdw}. For $n>1$, we will use the constraint on adjacent triples~\eqref{triplet_rel}:
\begin{equation}
c({\tt X baa Y})+c({\tt X bba Y})+c({\tt X bca Y}) = 0 = c({\tt X caa Y})+c({\tt X cba Y})+c({\tt X cca Y}) \,.
\label{eq:gentriples}
\end{equation}
The first equation follows from eq.~\eqref{triplet_rel} by the dihedral flip ${\tt a}\leftrightarrow {\tt b}$,
while the second equation follows from the first one by the dihedral flip ${\tt b}\leftrightarrow {\tt c}$.  Thus, eq.~\eqref{eq:gentriples} requires the sum of three coefficients to vanish.

Suppose we know that all strings ending in ${\tt ba}^n{\tt f}$ and ${\tt ca}^n{\tt f}$ have a vanishing coefficient.  Then in particular we know that all strings ending in ${\tt bba}^n{\tt f}$ and ${\tt bca}^n{\tt f}$ vanish.  By the first 
eq.~\eqref{eq:gentriples}, we know that all strings ending in ${\tt baa}^n{\tt f}$ vanish.  Similarly, using the second eq.~\eqref{eq:gentriples}, all strings ending in ${\tt caa}^n{\tt f}$ vanish. We have therefore proven that words ending in ${\tt ba}^{n+1}{\tt f}$ and ${\tt ca}^{n+1}{\tt f}$ have vanishing coefficients. By induction on $n$, all words ending in  ${\tt ba}^n{\tt f}$ and ${\tt ca}^n{\tt f}$ have vanishing coefficients, for any $n\geq 1$. Thus we have proven that the rules~\eqref{eq:generalized zero suffices appendix} can be deduced from antipodal duality \citep{Dixon:2021tdw} and the constraints from Ref.~\citep{Dixon:2022rse}.

\section{Counting allowed terms} \label{app:combinatorics}

The adjacency rules \ref{initialrule}--\ref{secondadjacency} from 
Sec.~\ref{sec:formfactor} indicate that nonzero elements must begin with ${\tt a},{\tt b}$ or ${\tt c}$, and end with ${\tt d,e}$ or ${\tt f}$. Also, they must not have adjacent ${\tt a}$ and ${\tt d}$, ${\tt b}$ and ${\tt e}$, or ${\tt c}$ and ${\tt f}$; nor adjacent ${\tt d}$ and ${\tt e}$, ${\tt d}$ and ${\tt f}$, or ${\tt e}$ and ${\tt f}$. In this appendix, we derive a formula for counting the number of elements satisfying these adjacency rules.  We also generalize the counting arguments to include zeroes related to multiple-final-entry relations, in particular eq.~\eqref{eq:generalized zero suffices appendix}.

We first consider the $6 \times 6$ transition matrix encoding the rules \ref{firstadjacency} and \ref{secondadjacency}:
\begin{equation}
    U=\begin{pmatrix}1&1&1&0&1&1\\ 1&1&1&1&0&1\\ 1&1&1&1 &1&0\\ 0&1&1&1&0&0\\1&0&1&0&1&0\\ 1&1&0&0&0&1\end{pmatrix},
\end{equation}
where entries of $U$ indicate the allowed sequences of two of the six letters: $U_{1,2}=1$ because ${\tt b}$ can follow ${\tt a}$, but $U_{1,4}=0$ because ${\tt d}$ cannot follow ${\tt a}$. Moreover, we consider the initial and final vectors $I=(1,1,1,0,0,0)$ and $F_1=(0,0,0,1,1,1)$, which encode rules \ref{initialrule} and \ref{finalrule}, respectively. 
The number of elements at weight $j$ that satisfy the four adjacency rules is 
\begin{equation}
\label{eq: counting zeros}
N_{\text{adj}}(j) = I U^{j-1} F_1^T .
\end{equation}
For loop order $L$ we require weight $j=2L$, i.e.~$N_{\text{adj}}(2L)$.  Explicit values of $N_{\text{adj}}(2L)$ for $L=1,\dots,8$ are given in the middle row of table~\ref{tab:elem_count1}, and again in the top row of table~\ref{tab:elem_count}. 

Because the adjacency rules~\ref{initialrule}--\ref{secondadjacency} respect dihedral symmetry, we can compress the transition matrix $U$ into a two-dimensional one, $\hat{U}$.  Effectively, we just track the number of allowed sequences at weight $j$ that end in an {\tt a} ($N_j^a$) versus those ending in a {\tt d} ($N_j^d$):
\begin{equation}
\label{eq:countingzeroshat}
N_{\text{adj}}(j) = N_j^d = \hat{I} \hat{U}^{j-1} \hat{F}_1^T , \qquad  N_j^a = \hat{I} \hat{U}^{j-1} \hat{A}^T ,
\end{equation}
where
\begin{equation}
    \hat{U}=\begin{pmatrix}3&2\\ 2&1\end{pmatrix}, \qquad \hat{I} = (3,0), \qquad \hat{F}_1 = (0,1), \qquad \hat{A} = (1,0).
    \label{eq:Uhat}
\end{equation}
It is easy to see that $N_j^a$ and $N_j^d$ obey the coupled recursion relations,
\begin{equation}
\begin{split}
N_j^a &= 3 N_{j-1}^a + 2 N_{j-1}^d \,, \\
N_j^d &= 2 N_{j-1}^a + N_{j-1}^d \,.
\end{split}
\label{eq:Njadrecursion}
\end{equation}
Combining these relations with the same ones for $j\to j-1$, one can obtain Fibonacci-like recursion relations for $N_j^a$ and $N_j^d$ individually.  For example,
\begin{equation}
N_j^d - 4 N_{j-1}^d = 2 N_{j-1}^a + (1-4) N_{j-1}^d = 2 ( 3 N_{j-2}^a + 2 N_{j-2}^d ) - 3 ( 2 N_{j-2}^a + N_{j-2}^d ) = N_{j-2}^d \,.
\end{equation}
Thus,
\begin{equation}
\begin{split}
N_j^a &= 4 N_{j-1}^a + N_{j-2}^a \,, \\
N_j^d &= 4 N_{j-1}^d + N_{j-2}^d \,.
\end{split}
\label{eq:NjadFibrecursion}
\end{equation}

The second eq.~\eqref{eq:NjadFibrecursion} is equivalent to
\begin{equation}
    N_{\text{adj}}(j) = 4 \, N_{\text{adj}}(j-1) + N_{\text{adj}}(j-2),
    \label{eq:FibNadj}
\end{equation}
for $j>2$, with the initial conditions $N_{\text{adj}}(1)=0$, $N_{\text{adj}}(2)=6$.  From eq.~\eqref{eq:FibNadj}, it is easy to see that the asymptotic growth rate of $N_{\text{adj}}(j)$ (as well as $N_j^a$) involves the solution to the quadratic equation $x^2=4x+1$,
\begin{equation}
    \frac{N_{\text{adj}}(j)}{N_{\text{adj}}(j-1)} \sim 2+\sqrt{5} = 4.236\ldots.
\end{equation}
The asymptotic growth rate of the number of adjacency-allowed terms at loop order $L$ is the square of this ratio, $(2+\sqrt{5})^2 \approx 17.94$.

Formula \eqref{eq:countingzeroshat} can be generalized to take into account
zeroes that might arise from the many linear relations among the last $k$ entries \citep{Dixon:2022rse}.
For example, the triple-final-entry conditions ($k=3$) 
include some zeroes that are not consequences of the pair adjacency relations, namely
\begin{equation}
c({\tt X baf}) = c({\tt Xcaf}) = 0\,,
\label{triplezero}
\end{equation}
plus dihedral images of these relations.  Eq.~\eqref{triplezero} is a special case of eq.\ \eqref{eq:generalized zero suffices appendix}.
Applying all of the $k=3$ relations, we find that there are 4 nonzero final triplets starting with ${\tt a}$ and 5 nonzero triplets starting with ${\tt d}$, or $F_3^a = 4$, $F_3^d = 5$.
\begin{table}[t]
    \small
    \centering
    \begin{tabular}{lccccccccc}
        \toprule
      $k$  & 1 & 2 & 3 & 4 & 5 & 6 & 7 & 8  \\
        \midrule
        $F_k^a$ & 0 & 2 & 4 & 18 & 76 & 322 & 1,364 & 5,778\\
        $F_k^d$ & 1 & 1 & 5 & 13 & 49 & 201 & 845 & 3,573\\
       \bottomrule
    \end{tabular}
    \caption{\small Number of nonzero $k$-final entries starting with $a$ and $d$, respectively.}
    \label{tab:F_k_count}
\end{table}
In general, if we know the vector
$\hat{F}_k \equiv (F_k^a,F_k^d)$, then we can compute the ``$k$-improved'' number of allowed nonzero elements, 
\begin{equation}
N(k,L) = \hat{I} \hat{U}^{2L-k} \hat{F}_k^T , \qquad \hbox{for $k\leq 2L$}.
\label{eq:NkL}
\end{equation}

We have used the final-entry relations from Ref.\ \citep{Dixon:2022rse} to compute $F_k^a$ and $F_k^d$ through $k=8$, with the results tabulated in table~\ref{tab:F_k_count}.  Interestingly, they also obey Fibonacci-like relations for $k>3$ through $k=8$:
\begin{equation}
\begin{split}
F_k^a &= 4 F_{k-1}^a + F_{k-2}^a \,,  \\
F_k^d &= 4 F_{k-1}^d + F_{k-2}^d - 8 \,. 
\end{split}
\label{eq:Fkad_recur}
\end{equation}
These relations predict $F_9^a = $24,476, $F_9^d = $15,129.  (This prediction could be tested using the antipodally-dual weight-9 hexagon function space on the parity-preserving surface.)

Now we will show that eqs.~\eqref{eq:Fkad_recur} follow from the {\it one-run zeroes hypothesis}.
This hypothesis is that the \emph{only} zeroes associated with multi-final-entry conditions are for ${\tt X ba\ldots af}$ and ${\tt Xca\ldots af}$ and their dihedral images, i.e.~eq.~\eqref{eq:generalized zero suffices}. Given the letter ${\tt d}$, any sub-word starting with ${\tt b}$ or ${\tt c}$, or ${\tt d}$ can sit behind it.  Thus,
\begin{equation}
F_k^d = 2 F_{k-1}^a + F_{k-1}^d \,.
\label{eq:recurFkd}
\end{equation}
On the other hand, we need to divide $F_k^a$ into $F_k^a-2$ and 2, where the latter 2 is associated with the two sub-words ${\tt a\ldots ae}$ and ${\tt a\ldots af}$.  Due to eq.~\eqref{eq:generalized zero suffices}, these latter sub-words each generate one new sub-word at the next value of $k$, whereas the $F_k^a-2$ can have all 3 of ${\tt a,b,c}$ prepended to them.  Thus,
\begin{align}
F_k^a &= 3 (F_{k-1}^a - 2) + 2 + 2 F_{k-1}^d 
\label{eq:recurFka}\\
  &= 3 F_{k-1}^a + 2 F_{k-1}^d - 4 \,, \nonumber
\end{align}
where the last term on the first line ($2 F_{k-1}^d$) counts sub-words of the form ${\tt aeZ}$ and ${\tt afZ}$, for some sub-word ${\tt Z}$.
It is straightforward to combine eqs.~(\ref{eq:recurFkd}) and (\ref{eq:recurFka}) with the same equations with $k \to k-1$, in order to derive eqs.~\eqref{eq:Fkad_recur}.
For example, 
\begin{align}
F_k^a - 4 F_{k-1}^a &= - F_{k-1}^a + 2 F_{k-1}^d - 4 \nonumber\\
  &= - ( 3 F_{k-2}^a + 2 F_{k-2}^d - 4 ) + 2 (2 F_{k-2}^a + F_{k-2}^d ) - 4\nonumber\\
  &= F_{k-2}^a \,.
  \label{eq:deriveFka}
\end{align}

Notice that the form of eqs.~\eqref{eq:Fkad_recur} is very similar to eqs.~\eqref{eq:NjadFibrecursion}.  At large $k$, the constant term $-8$ is negligible, and the asymptotic growth rate for $F_k^a$ and $F_k^d$ is the same as it is for $N_j^a$ and $N_j^d$, namely $2+\sqrt{5}$.

Using \eqn{eq:NkL} and table~\ref{tab:F_k_count}, we give the number of $k$-improved allowed nonzero elements in table~\ref{tab:k_allowed}.  The number decreases with $k$, but the decrease from $k=7$ to $k=8$ (say) is not very great, and the decrease stops once $k=2L$. The $k=\infty$ line assumes the one-run zeroes hypothesis. In the last two lines of the table, we give the fractions of possibly-nonzero entries that are actually nonzero, for $k=8$ and for $k=\infty$. There is very little difference between $k=8$ and $k=\infty$. For $L\geq4$, this fraction increases monotonically with $L$, reaching $93\%$ by $L=8$.  Thus the majority of the zeroes can be understood to stem from the adjacency relations together with the $k$-final-entry relations.  On the other hand, the absolute number of ``unexplained'' zeroes, $N(8,L)-{\rm actual}$, still increases quickly with $L$.

\begin{table}[t]
    \small
    \centering
    \begin{tabular}{lccccccccc}
        \toprule
      $L$  & 1 & 2 & 3 & 4 & 5 & 6 & 7 & 8  \\
        \midrule
        $k=3$ & 6 & 66 & 1,170 & 20,994 & 376,722 & 6,760,002 & 121,303,314 & 2,176,699,650 \\
        $k=4$ & 6 & 54 & 1,014 & 18,198 & 326,550 & 5,859,702 & 105,148,086 & 1,886,805,846\\
        $k=5$ & 6 & 54 & 978 & 17,538 & 314,706 & 5,647,170 & 101,334,354 & 1,818,371,202\\ 
        $k=6$ & 6 & 54 & 966 & 17,382 & 311,910 & 5,596,998 & 100,434,054 & 1,802,215,974\\
        $k=7$ & 6 & 54 & 966 & 17,346 & 311,250 & 5,585,154 & 100,221,522 & 1,798,402,242\\
         $k=8$ & 6 & 54 & 966 & 17,334 & 311,094 & 5,582,358 & 100,171,350 & 1,797,501,942\\   
         $k=\infty$ & 6 & 54 & 966 & 17,334 & 311,046 & 5,581,494 & 100,155,846 & 1,797,223,734\\
         \bottomrule
   \hbox{actual} & 6 & 12 & 636 & 11,208 & 263,880 & 4,916,466 & 
          92,954,568 & 1,671,656,292\\
          \bottomrule
    $N(8,L)-\hbox{actual}$ & 0 & 42 & 330 & 6,126 & 47,214 & 665,892 & 7,216,782 & 125,845,650\\ 
    ${\rm actual}/N(8,L)$ & 1. & 0.222 & 0.658 & 0.6466 & 0.8482 & 
          0.8807 & 0.9279 & 0.9299\\
    ${\rm actual}/N(\infty,L)$ & 1. & 0.222 & 0.658 & 0.6466 & 0.8484 & 
          0.8809 & 0.9281 & 0.9301\\
       \bottomrule
    \end{tabular}
    \caption{\small Number of nonzero elements at $L$ loop that are permitted by the $k$-final-entry relations and the adjacency relations, in comparison with the actual number of nonzero elements. The $k=\infty$ line assumes the one-run zeroes hypothesis.}
    \label{tab:k_allowed}
\end{table}

If we subtract the $k=\infty$ line of table~\ref{tab:k_allowed} from the ``adjacency-allowed'' line of table~\ref{tab:elem_count}, we obtain the ``suffix-forbidden'' line of table~\ref{tab:elem_count}.  Next we need to count the number of prefix-forbidden elements that are left.  This number is just $6\,F_{2L-2}^d$, because 6 types of strings with a {\tt d}-type letter in the third symbol slot have to be removed, and the number of each type that is allowed by the suffix rules is just $F_{k}^d$ for $k=2L-2$. Subtracting the suffix- and prefix-forbidden elements from the adjacency-allowed elements gives the remaining allowed elements in table~\ref{tab:elem_count}.  Then it is just a matter of comparing this number with the actual number of nonzero terms in the symbol~\citep{Dixon:2021tdw}.

At $L=2$, there are 42 unexplained zeroes which fall into seven 6-orbits under the dihedral group.  Representatives of the seven orbits are
\begin{equation}
{\tt abdd},\ {\tt aecd},\ {\tt bbdd},\ {\tt bcdd},\ 
{\tt bdbd},\ {\tt bdcd},\ {\tt bfbd}.
\end{equation}
The first of the seven orbits can be explained because {\tt abdd} doesn't appear in the 48-dimensional space of allowed first-four entries. This is part of the prefix rule~\eqref{eq: prefix zero}.  But the other six vanishing 6-orbits at two loops do not seem to have a simple explanation.  There is still considerably more to understand about the unexplained zeroes at high loop order.

In summary, in this appendix we computed the number of allowed terms in the symbol due to the adjacency relations and the $k$-final-entry conditions, in particular the suffix rules~\eqref{eq:generalized zero suffices}.  We also counted the number of terms forbidden by the prefix rule~\eqref{eq: prefix zero}.  We found that the auxiliary sequences $N_j^a$, $N_j^d$, $F_k^a$, $F_k^d$ all obey Fibonacci-like sequences with a growth rate of $2+\sqrt{5}$ per weight, or $(2+\sqrt{5})^2$ per loop. This rate originates as one of the eigenvalues of the matrix $\hat{U}$ in eq.~\eqref{eq:Uhat}. This rate will control $N(k,L)$ for any fixed $k$.  It will also control the improved number of allowed terms that takes into account the prefix rule~\eqref{eq: prefix zero}. From the last line of table~\ref{tab:elem_count}, the ratio of the actual number of terms to the remaining allowed terms is over 99.8\% by $L=8$. Hence it is very likely that the actual number of nonzero terms in the symbol also has an asymptotic growth rate per loop of $(2+\sqrt{5})^2\approx 17.94$.

\section{Bootstrapping using all-loop sequences}
\label{app: bootstrapping}

In this appendix, we analyze the power of all-loop sequences and empirical linear relations, demonstrating their potential to improve the symbol bootstrap approach.

Suppose we wish to train an AI model to predict loop orders that are (mostly) not yet known.  Suppose that the AI model can learn all of the known relations in 
Sec.~\ref{sec:formfactor} that include dihedral symmetry~\eqref{dihedralsymm}, integrability~\eqref{commute_left}--\eqref{eq:14terms}, the adjacency rules, and the $k$-final-entry conditions.  The multi-final-entry conditions are related by antipodal duality to the hexagon function space, so we suppose we know what they are for any $k$. 

Let's also suppose that we can enforce a branch-cut condition and a constraint on the $L^{\rm th}$ discontinuity, discussed below.  In addition, as described at the end of Sec.~\ref{sec:rays}, at $L$ loops we should have available all-loop results for all sequences ending in $(L+1)$ {\tt f}'s.  Is that enough to fix the form factor completely at a given loop order $L$?  

When the form-factor function space is constructed, it is natural to include a branch-cut condition.  This condition is only ``semi-local'', by which we mean that it involves an increasingly large number of terms in the $\{a,b,c,d,e,f\}$ alphabet at higher loop orders.  The branch-cut condition states that if there is an {\tt f} in the symbol, then in the kinematical region where {\tt f} vanishes ($w\to1$, $u,v\to0$), the terms preceding the {\tt f} should vanish. One sets
\begin{equation}
a = \sqrt{\frac{u}{v}} \,, \quad 
b = \sqrt{\frac{v}{u}} \,, \quad
c = \sqrt{\frac{1}{uv}} \,, \quad
d = \frac{1}{u} \,, \quad
e = \frac{1}{v} \,,
\end{equation}
or equivalently makes the substitutions
\begin{equation} 
b \rightarrow \frac{1}{a} \,, \quad
d \rightarrow \frac{c}{a} \,, \quad
e \rightarrow a\times c \,.
\end{equation}
One then collects the terms with symbol letters now only belonging to {\tt a,c},
and requires those terms preceding each {\tt f} to vanish.  We also impose the empirical multi-initial-entry relations mentioned at the end of Sec.~\ref{sec:formfactor}.

There is also a known constraint on the $L^{\rm th}$ discontinuity of the form factor~\citep{Dixon:2022rse}.  The $L^{\rm th}$ discontinuity is computed by clipping $L$ letters off the front of the symbol.  However, the letter that is supposed to be clipped is $w$, which corresponds to the operation
\begin{equation}
  {\rm Disc}_w
  = \frac{1}{2} \Bigl[ - {\rm Disc}_a - {\rm Disc}_b + {\rm Disc}_c \Bigr]
       - {\rm Disc}_f \,.
\end{equation}
This constraint is also ``semi-local'', involving more and more terms at higher loops.

\renewcommand{\arraystretch}{1.25}
\begin{table}[!t]
\centering
\begin{tabular}[t]{l c c c c c}
\hline\hline
$n$ final entries & fe$_4$ & fe$_5$ & fe$_6$ & fe$_7$ & fe$_8$
\\\hline\hline
fe$_k$ only               &    38  &  18    &   12  &   10 &   9
\\\hline
+ {\tt f}$^6$             &    28  &   9    &    4  &   4  &   4
\\\hline
+ disc              &    29  &  12    &    7  &   5  &   4
\\\hline
+ {\tt f}$^6$ + disc      &    19  &   4    &    1  &   1  &   1
\\\hline
\end{tabular}
\caption{Unknown parameters remaining at $L=5$ loops. We always impose dihedral symmetry, multi-initial-entry constraints, integrability and adjacency.  In the top line, we impose the constraints on the $k$ final entries, $k=5,6,7,8$.  In the next line, we add the all-loop constraint that we know all sequences ending in 6 {\tt f}'s.   In the line after that, we apply the $L^{\rm th}$ discontinuity constraint (but not {\tt f}$^6$).  In the final line, we add both the {\tt f}$^6$ and 
$L^{\rm th}$ discontinuity constraints, arriving at a single unknown parameter.}
\label{tab:Leq5}
\end{table}

At all loop orders from 3 to 7, we solved the linear equations resulting from these constraints, and counted how many unknown parameters were left.  Table~\ref{tab:Leq5} shows the results at 5 loops.  Even imposing up to 8-final-entry conditions, there are still 9 parameters left.  If we impose knowledge of the elements ending in 6 {\tt f}'s, the number of parameters is reduced to 4.  If we instead impose the $L^{\rm th}$ discontinuity constraint, it also leaves 4 parameters.  But if we impose both {\tt f}$^6$ {\it and} the $L^{\rm th}$ discontinuity constraint, there is a single parameter left.

Remarkably, for $L=3,4,5,6,7$ there is a single parameter left as well, after imposing the same set of constraints as in table~\ref{tab:Leq5}, including fe$_k$ for $k\leq8$, {\tt f}$^{L+1}$, and the $L^{\rm th}$ discontinuity. At $L=2$, there is no parameter left, i.e.~the answer is uniquely determined.   At $L=8$, we have not yet performed the analysis, but we conjecture it that there is exactly one parameter left also for $L\geq 8$.   Also somewhat remarkably, the ``ambiguity function'' multiplying the one parameter at $L=3,4,5,6,7$ vanishes in the strict collinear limit.  It can be fixed by the FFOPE, using the first nontrivial (leading-logarithmic) contribution to the FFOPE.

In conclusion, the all-loop information about sequences ending in $(L+1)$ {\tt f}'s is quite powerful, and quite complementary to the $L^{\rm th}$ discontinuity constraint, leaving behind, at symbol level, a single unknown parameter to fix using the FFOPE information.  Perhaps another all-loop sequence can be found that fixes it, without having to resort to the FFOPE at all.  It will also be very interesting to see to what extent an AI model can learn all the types of constraints used in this analysis (or alternative ones).

\providecommand{\href}[2]{#2}\begingroup\raggedright\endgroup


\begin{thebibliography}{10}

\bibitem{Travaglini:2022uwo}
G.~Travaglini et~al., \emph{{The SAGEX review on scattering amplitudes}},
  \href{https://doi.org/10.1088/1751-8121/ac8380}{\emph{J. Phys. A} {\bfseries
  55} (2022) 443001} [\href{https://arxiv.org/abs/2203.13011}{{\ttfamily
  2203.13011}}].

\bibitem{Wilczek:1977zn}
F.~Wilczek, \emph{{Decays of Heavy Vector Mesons Into Higgs Particles}},
  \href{https://doi.org/10.1103/PhysRevLett.39.1304}{\emph{Phys. Rev. Lett.}
  {\bfseries 39} (1977) 1304}.

\bibitem{Shifman:1978zn}
M.~A. Shifman, A.~I. Vainshtein and V.~I. Zakharov, \emph{{Remarks on Higgs
  Boson Interactions with Nucleons}},
  \href{https://doi.org/10.1016/0370-2693(78)90481-1}{\emph{Phys. Lett. B}
  {\bfseries 78} (1978) 443}.

\bibitem{Gehrmann:2011aa}
T.~Gehrmann, M.~Jaquier, E.~W.~N. Glover and A.~Koukoutsakis, \emph{{Two-Loop
  QCD Corrections to the Helicity Amplitudes for $H \to$ 3 partons}},
  \href{https://doi.org/10.1007/JHEP02(2012)056}{\emph{JHEP} {\bfseries 02}
  (2012) 056} [\href{https://arxiv.org/abs/1112.3554}{{\ttfamily 1112.3554}}].

\bibitem{Brandhuber:2012vm}
A.~Brandhuber, G.~Travaglini and G.~Yang, \emph{{Analytic two-loop form factors
  in N=4 SYM}}, \href{https://doi.org/10.1007/JHEP05(2012)082}{\emph{JHEP}
  {\bfseries 05} (2012) 082} [\href{https://arxiv.org/abs/1201.4170}{{\ttfamily
  1201.4170}}].

\bibitem{Dixon:2021tdw}
L.~J. Dixon, {\"{O}}.~{G\"{u}rdo\u{g}an}, A.~J. McLeod and M.~Wilhelm,
  \emph{{Folding Amplitudes into Form Factors: An Antipodal Duality}},
  \href{https://doi.org/10.1103/PhysRevLett.128.111602}{\emph{Phys. Rev. Lett.}
  {\bfseries 128} (2022) 111602}
  [\href{https://arxiv.org/abs/2112.06243}{{\ttfamily 2112.06243}}].

\bibitem{Gehrmann:2000zt}
T.~Gehrmann and E.~Remiddi, \emph{{Two loop master integrals for $\gamma^*
  \rightarrow$ 3 jets: The Planar topologies}},
  \href{https://doi.org/10.1016/S0550-3213(01)00057-8}{\emph{Nucl. Phys. B}
  {\bfseries 601} (2001) 248}
  [\href{https://arxiv.org/abs/hep-ph/0008287}{{\ttfamily hep-ph/0008287}}].

\bibitem{Dixon:2020bbt}
L.~J. Dixon, A.~J. McLeod and M.~Wilhelm, \emph{{A Three-Point Form Factor
  Through Five Loops}},
  \href{https://doi.org/10.1007/JHEP04(2021)147}{\emph{JHEP} {\bfseries 04}
  (2021) 147} [\href{https://arxiv.org/abs/2012.12286}{{\ttfamily
  2012.12286}}].

\bibitem{Dixon:2022rse}
L.~J. Dixon, {\"{O}}.~{G\"{u}rdo\u{g}an}, A.~J. McLeod and M.~Wilhelm,
  \emph{{Bootstrapping a stress-tensor form factor through eight loops}},
  \href{https://doi.org/10.1007/JHEP07(2022)153}{\emph{JHEP} {\bfseries 07}
  (2022) 153} [\href{https://arxiv.org/abs/2204.11901}{{\ttfamily
  2204.11901}}].

\bibitem{Beisert:2010jr}
N.~Beisert et~al., \emph{{Review of AdS/CFT Integrability: An Overview}},
  \href{https://doi.org/10.1007/s11005-011-0529-2}{\emph{Lett. Math. Phys.}
  {\bfseries 99} (2012) 3} [\href{https://arxiv.org/abs/1012.3982}{{\ttfamily
  1012.3982}}].

\bibitem{Basso:2013vsa}
B.~Basso, A.~Sever and P.~Vieira, \emph{{Spacetime and Flux Tube S-Matrices at
  Finite Coupling for N=4 Supersymmetric Yang-Mills Theory}},
  \href{https://doi.org/10.1103/PhysRevLett.111.091602}{\emph{Phys. Rev. Lett.}
  {\bfseries 111} (2013) 091602}
  [\href{https://arxiv.org/abs/1303.1396}{{\ttfamily 1303.1396}}].

\bibitem{Sever:2020jjx}
A.~Sever, A.~G. Tumanov and M.~Wilhelm, \emph{{Operator Product Expansion for
  Form Factors}},
  \href{https://doi.org/10.1103/PhysRevLett.126.031602}{\emph{Phys. Rev. Lett.}
  {\bfseries 126} (2021) 031602}
  [\href{https://arxiv.org/abs/2009.11297}{{\ttfamily 2009.11297}}].

\bibitem{Sever:2021nsq}
A.~Sever, A.~G. Tumanov and M.~Wilhelm, \emph{{An Operator Product Expansion
  for Form Factors II. Born level}},
  \href{https://doi.org/10.1007/JHEP10(2021)071}{\emph{JHEP} {\bfseries 10}
  (2021) 071} [\href{https://arxiv.org/abs/2105.13367}{{\ttfamily
  2105.13367}}].

\bibitem{Sever:2021xga}
A.~Sever, A.~G. Tumanov and M.~Wilhelm, \emph{{An Operator Product Expansion
  for Form Factors III. Finite Coupling and Multi-Particle Contributions}},
  \href{https://doi.org/10.1007/JHEP03(2022)128}{\emph{JHEP} {\bfseries 03}
  (2022) 128} [\href{https://arxiv.org/abs/2112.10569}{{\ttfamily
  2112.10569}}].

\bibitem{Dixon:2011pw}
L.~J. Dixon, J.~M. Drummond and J.~M. Henn, \emph{{Bootstrapping the three-loop
  hexagon}}, \href{https://doi.org/10.1007/JHEP11(2011)023}{\emph{JHEP}
  {\bfseries 11} (2011) 023} [\href{https://arxiv.org/abs/1108.4461}{{\ttfamily
  1108.4461}}].

\bibitem{Caron-Huot:2016owq}
S.~Caron-Huot, L.~J. Dixon, A.~McLeod and M.~von Hippel, \emph{{Bootstrapping a
  Five-Loop Amplitude Using Steinmann Relations}},
  \href{https://doi.org/10.1103/PhysRevLett.117.241601}{\emph{Phys. Rev. Lett.}
  {\bfseries 117} (2016) 241601}
  [\href{https://arxiv.org/abs/1609.00669}{{\ttfamily 1609.00669}}].

\bibitem{Caron-Huot:2019vjl}
S.~Caron-Huot, L.~J. Dixon, F.~Dulat, M.~von Hippel, A.~J. McLeod and
  G.~Papathanasiou, \emph{{Six-Gluon amplitudes in planar $ \mathcal{N} $ = 4
  super-Yang-Mills theory at six and seven loops}},
  \href{https://doi.org/10.1007/JHEP08(2019)016}{\emph{JHEP} {\bfseries 08}
  (2019) 016} [\href{https://arxiv.org/abs/1903.10890}{{\ttfamily
  1903.10890}}].

\bibitem{Caron-Huot:2020bkp}
S.~Caron-Huot, L.~J. Dixon, J.~M. Drummond, F.~Dulat, J.~Foster,
  O.~G\"urdo\u{g}an et~al., \emph{{The Steinmann Cluster Bootstrap for $N$ = 4
  Super Yang-Mills Amplitudes}},
  \href{https://doi.org/10.22323/1.376.0003}{\emph{PoS} {\bfseries CORFU2019}
  (2020) 003} [\href{https://arxiv.org/abs/2005.06735}{{\ttfamily
  2005.06735}}].

\bibitem{Goncharov:2010jf}
A.~B. Goncharov, M.~Spradlin, C.~Vergu and A.~Volovich, \emph{{Classical
  Polylogarithms for Amplitudes and Wilson Loops}},
  \href{https://doi.org/10.1103/PhysRevLett.105.151605}{\emph{Phys. Rev. Lett.}
  {\bfseries 105} (2010) 151605}
  [\href{https://arxiv.org/abs/1006.5703}{{\ttfamily 1006.5703}}].

\bibitem{Goncharov:2001iea}
A.~B. Goncharov, \emph{{Multiple polylogarithms and mixed Tate motives}},
  \href{https://arxiv.org/abs/math/0103059}{{\ttfamily math/0103059}}.

\bibitem{Cai:2024znx}
T.~Cai, G.~W. Merz, F.~Charton, N.~Nolte, M.~Wilhelm, K.~Cranmer et~al.,
  \emph{{Transforming the bootstrap: using transformers to compute scattering
  amplitudes in planar $\mathcal{N} = 4$ super Yang\textendash{}Mills theory}},
  \href{https://doi.org/10.1088/2632-2153/ad743e}{\emph{Mach. Learn. Sci.
  Tech.} {\bfseries 5} (2024) 035073}
  [\href{https://arxiv.org/abs/2405.06107}{{\ttfamily 2405.06107}}].

\bibitem{Vaswani:2017lxt}
A.~Vaswani, N.~Shazeer, N.~Parmar, J.~Uszkoreit, L.~Jones, A.~N. Gomez et~al.,
  \emph{{Attention Is All You Need}},  in \emph{{31st International Conference
  on Neural Information Processing Systems}}, 2017,
  \href{https://arxiv.org/abs/1706.03762}{{\ttfamily 1706.03762}}.

\bibitem{Brandhuber:2011tv}
A.~Brandhuber, {\"{O}}.~{G\"{u}rdo\u{g}an}, R.~Mooney, G.~Travaglini and
  G.~Yang, \emph{{Harmony of Super Form Factors}},
  \href{https://doi.org/10.1007/JHEP10(2011)046}{\emph{JHEP} {\bfseries 10}
  (2011) 046} [\href{https://arxiv.org/abs/1107.5067}{{\ttfamily 1107.5067}}].

\bibitem{Bern:2005iz}
Z.~Bern, L.~J. Dixon and V.~A. Smirnov, \emph{{Iteration of planar amplitudes
  in maximally supersymmetric Yang-Mills theory at three loops and beyond}},
  \href{https://doi.org/10.1103/PhysRevD.72.085001}{\emph{Phys. Rev. D}
  {\bfseries 72} (2005) 085001}
  [\href{https://arxiv.org/abs/hep-th/0505205}{{\ttfamily hep-th/0505205}}].

\bibitem{Goncharov:1998kja}
A.~B. Goncharov, \emph{{Multiple polylogarithms, cyclotomy and modular
  complexes}}, \href{https://doi.org/10.4310/MRL.1998.v5.n4.a7}{\emph{Math.
  Res. Lett.} {\bfseries 5} (1998) 497}
  [\href{https://arxiv.org/abs/1105.2076}{{\ttfamily 1105.2076}}].

\bibitem{Duhr:2014woa}
C.~Duhr, \emph{{Mathematical aspects of scattering amplitudes}},  in
  \emph{{Theoretical Advanced Study Institute in Elementary Particle Physics}:
  {Journeys Through the Precision Frontier: Amplitudes for Colliders}},
  pp.~419--476, 2015, \href{https://arxiv.org/abs/1411.7538}{{\ttfamily
  1411.7538}}, \href{https://doi.org/10.1142/9789814678766_0010}{DOI}.

\bibitem{Dixon:2013eka}
L.~J. Dixon, J.~M. Drummond, M.~von Hippel and J.~Pennington, \emph{{Hexagon
  functions and the three-loop remainder function}},
  \href{https://doi.org/10.1007/JHEP12(2013)049}{\emph{JHEP} {\bfseries 12}
  (2013) 049} [\href{https://arxiv.org/abs/1308.2276}{{\ttfamily 1308.2276}}].

\bibitem{Caron-Huot:2019bsq}
S.~Caron-Huot, L.~J. Dixon, F.~Dulat, M.~Von~Hippel, A.~J. McLeod and
  G.~Papathanasiou, \emph{{The Cosmic Galois Group and Extended Steinmann
  Relations for Planar $\mathcal{N} = 4$ SYM Amplitudes}},
  \href{https://doi.org/10.1007/JHEP09(2019)061}{\emph{JHEP} {\bfseries 09}
  (2019) 061} [\href{https://arxiv.org/abs/1906.07116}{{\ttfamily
  1906.07116}}].

\bibitem{OEISA052737}
{N. J. A.~Sloane~et~al.}, ``Online encyclopedia of integer sequences.''
\newblock \url{https://oeis.org/A052737}.

\bibitem{Basso:2024hlx}
B.~Basso, L.~J. Dixon and A.~G. Tumanov, \emph{{The Three-Point Form Factor of
  $\textrm{Tr}\,\phi^3$ to Six Loops}},
\href{https://doi.org/10.1007/JHEP02(2025)034}{\emph{JHEP} {\bfseries 02}
  (2025) 034} [\href{https://arxiv.org/abs/2410.22402}{{\ttfamily 2410.22402}}].

\bibitem{Henn:2024pki}
J.~M. Henn, J.~Lim and W.~J. Torres~Bobadilla, \emph{{Analytic evaluation of
  the three-loop three-point form factor of $\operatorname{tr}\phi^3$ in
  $\mathcal{N}=4$ sYM}},
\href{https://doi.org/10.1007/JHEP02(2025)085}{\emph{JHEP} {\bfseries 02}
  (2025) 085} [\href{https://arxiv.org/abs/2410.22465}{{\ttfamily
  2410.22465}}].

\end{thebibliography}
\end{document}